\documentclass[aps, prl, 10pt, twocolumn,superscriptaddress]{revtex4-1}

\usepackage[utf8]{inputenc}
\usepackage{epsfig}
\usepackage{graphicx}
\usepackage{float}
\usepackage{dcolumn}
\usepackage{xcolor}
\usepackage{bm}
\usepackage{amsmath,bbm}
\usepackage{gensymb}

\DeclareUnicodeCharacter{2212}{-}
\begin{document}

\title{Parity-preserving and magnetic field resilient superconductivity in indium antimonide nanowires with tin shells}

\author{M. Pendharkar}
\affiliation{Electrical and Computer Engineering, University of California Santa Barbara, Santa Barbara, CA 93106, USA}
\author{B. Zhang}
\affiliation{Department of Physics and Astronomy, University of Pittsburgh, Pittsburgh, PA 15260, USA} 
\author{H. Wu}
\affiliation{Department of Physics and Astronomy, University of Pittsburgh, Pittsburgh, PA 15260, USA} 
\author{A. Zarassi}
\affiliation{Department of Physics and Astronomy, University of Pittsburgh, Pittsburgh, PA 15260, USA} 
\author{P. Zhang}
\affiliation{Department of Physics and Astronomy, University of Pittsburgh, Pittsburgh, PA 15260, USA} 
\author{C.P. Dempsey}
\affiliation{Electrical and Computer Engineering, University of California Santa Barbara, Santa Barbara, CA 93106, USA}
\author{\\J.S. Lee}
\affiliation{California NanoSystems Institute, University of California Santa Barbara, Santa Barbara, CA 93106, USA}
\author{S.D. Harrington}
\affiliation{Materials Department, University of California Santa Barbara, Santa Barbara, CA 93106, USA}
\author{G. Badawy}
\affiliation{Eindhoven University of Technology, 5600 MB, Eindhoven, The Netherlands}
\author{S. Gazibegovic}
\affiliation{Eindhoven University of Technology, 5600 MB, Eindhoven, The Netherlands}
\author{J. Jung}
\affiliation{Eindhoven University of Technology, 5600 MB, Eindhoven, The Netherlands}
\author{A.-H. Chen}
\affiliation{Univ. Grenoble Alpes, CNRS, Grenoble INP, Institut N\'eel, 38000 Grenoble, France}
\author{\\M.A. Verheijen}
\affiliation{Eindhoven University of Technology, 5600 MB, Eindhoven, The Netherlands} 
\author{M. Hocevar}
\affiliation{Univ. Grenoble Alpes, CNRS, Grenoble INP, Institut N\'eel, 38000 Grenoble, France}
\author{E.P.A.M. Bakkers}
\affiliation{Eindhoven University of Technology, 5600 MB, Eindhoven, The Netherlands} 
\author{C.J. Palmstr\o m}
\affiliation{Electrical and Computer Engineering, University of California Santa Barbara, Santa Barbara, CA 93106, USA}
\affiliation{California NanoSystems Institute, University of California Santa Barbara, Santa Barbara, CA 93106, USA}
\affiliation{Materials Department, University of California Santa Barbara, Santa Barbara, CA 93106, USA}
\author{S.M. Frolov}
\email{frolovsm@pitt.edu}
\affiliation{Department of Physics and Astronomy, University of Pittsburgh, Pittsburgh, PA 15260, USA} 

\date{\today}

\begin{abstract}
We study bottom-up grown semiconductor indium antimonide nanowires that are coated with shells of tin. The shells are uniform in thickness. The interface between Sn and InSb is abrupt and without interdiffusion. Devices for transport are prepared by in-situ shadowing of nanowires using nearby nanowires as well as flakes, resulting in etch-free junctions. Tin is found to induce a hard superconducting gap in the range $600-700~\mu$eV. Superconductivity persists up to 4~T in magnetic field. A small island of InSb/Sn exhibits the coveted two-electron charging effect, a hallmark of charge parity stability. The findings open avenues for superconducting and topological quantum circuits based on new superconductor-semiconductor combinations.
\end{abstract}

\maketitle

As we enter the era of intermediate-scale quantum circuits~\cite{Arute2019,Figgatt2019},
materials considerations come into renewed focus through their impact on quantum gate fidelity.  The most successful solid state approaches rely either on superconductors~\cite{doi:10.1063/1.5089550}, or on semiconductors~\cite{kloeffel2013prospects}, with the future topological platform to require a hybrid of both~\cite{lutchynnrm2018}. The search continues for the ultimate material capable of rendering moot the issue of intrinsic decoherence. In this context, the push for qubits based on Majorana zero modes that are expected to be topologically immune to decoherence~\cite{Hyart2013,Karzig2017,Plugge2017, stenger2019braiding} has facilitated the introduction of high quality interfaces between superconducting metals and low-dimensional semiconductors~\cite{Krogstrup2015, Shabani2016, gazibegovic2017epitaxy, fornieri2019evidence}.

Only a few superconductors were explored for Majorana qubits, most notably aluminum which is also the material of choice for transmon quantum processors~\cite{Arute2019}. Among advantages of aluminum are self-limiting native oxide and hard induced gap in proximate semiconductors~\cite{Krogstrup2015,Chang2015,Gul2017,gazibegovic2017epitaxy}. Due to this, aluminum is widely known to exhibit 2e charging in small islands, a crucial basic property that makes it a low-decoherence superconductor~\cite{Geerligs1990,TuominenPRL92,LafargePRL93,EilesPRL93, Albrecht2016, Shen2018, carrad2019shadow}. Among disadvantages of aluminum are a relatively small superconducting gap equivalent to 1~K, and a low critical magnetic field.
This confines quantum computing to ultra-low temperatures and even further constrains how we design future topological qubits which will require a precise balance of several energy scales~\cite{pan2019zero}.  

Here we present induced superconductivity in InSb nanowires~\cite{badawy2019high,PlissardNanoLett12} with Sn shells. InSb is the highest electron mobility group III-V semiconductor with  strong spin-orbit coupling~\cite{nadj-pergePRL12} and large Land$\acute{e}$ g-factors in the conduction band~\cite{Nilsson2009}. These are the primary ingredients of the Majorana recipe~\cite{Lutchyn2010,Oreg2010}, making InSb an optimal material for the investigation of induced topological superconductivity~\cite{Mourik2012, Deng2012, Chen2017,GulNatNano18, ZhangNature2018}. 

We find that when InSb nanowires are coupled to tin they exhibit hard induced superconducting gap up to 700~$\mu$eV. Superconductivity persists to a significant magnetic field, up to 4 Tesla for 15~nm thick Sn shells. Most importantly, islands of tin do exhibit 2e-periodic charging patterns. This effect is a landmark requirement for topological quantum computing as well as for transmon qubits, as it is a pre-requisite for long quasiparticle stability times. Our results indicate that there are more superconductors compatible with quantum computing, opening future directions for heterostructure tailoring and ultimately for high fidelity quantum circuits.

\begin{figure*}
    \centering
    \includegraphics[width=\textwidth]{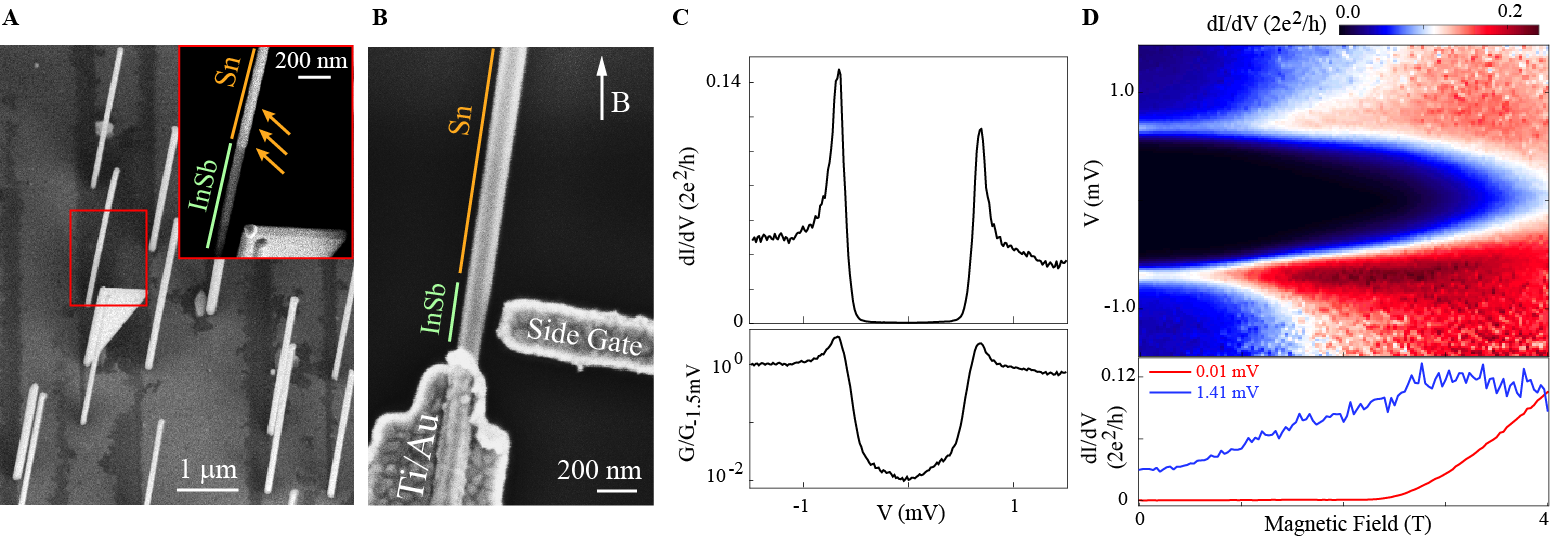}
    \caption{(A) Scanning electron micrograph (SEM) showing a triangular InSb flake that stood in the path of a beam of Sn atoms, shadowing the InSb nanowire standing behind. Dark streaks on the InSb substrate are also due to shadowing of the Sn beam by nanowires and flakes. The inset shows the direction of Sn beam and indicates the shadowed and exposed segments of the nanowire. (B) SEM of an N-S device (device A) with a flake-shadowed InSb/Sn nanowire and Ti/Au contacts and a side gate. Magnetic field is applied vertically. (C) Zero magnetic field tunneling conductance spectrum of device A in linear scale (top) and logarithmic scale (bottom), V$_\textrm{BG}=7.5$~V, V$_\textrm{SG}=-0.4$~V. (D) Magnetic field evolution of the spectrum in panel (C). Bottom part contain linecuts along dashed lines in the top part. Line traces are at low and high $V$.
    }
    \label{fig1}
\end{figure*}

Our first goal is to investigate electron tunneling into Sn through InSb in the normal metal-superconductor (N-S) configuration. For this we need a nanowire with only one end covered by tin. We use the uncovered end to define a tunneling barrier and an N-contact. In order to avoid damage to InSb that results from etching away part of the Sn shell, we employ in-situ nanoscale shadowing~\cite{gazibegovic2017epitaxy,carrad2019shadow}. In this case it is an InSb flake, standing in front of the nanowire, that shadows the bottom of the wire during deposition of Sn in ultra-high vacuum~\cite{gazibegovic2019bottom} (Fig.~\ref{fig1}A). 

To prepare an N-S device the tin-coated nanowire is positioned onto a doped Si/$\mathrm{SiO_x}$ substrate which is used as a back gate (BG) (Fig.~\ref{fig1}B). A side gate (SG) is used to define and tune the tunneling barrier near the edge of the tin-free segment. The tunneling spectrum reveals a two-orders-of-magnitude suppression in conductance around zero bias (Fig.~\ref{fig1}C). This so-called hard gap indicates the elimination of decoherence pathways due to disorder and spurious subgap states. A superconducting tunneling peak is at $\pm 680~\mu$eV which is comparable to the gap of tin. In magnetic field, the hard gap is found to persist beyond 2 Tesla, with the gap ``softening" at higher fields and fully closing around 4 Tesla (Fig.~\ref{fig1}D). Magnetic field resilience is an indicator of a thin uniform shell.
It is another advantage of Sn since topological, spin and some superconducting qubits operate at high magnetic fields. Supporting Materials contain extended data images and source files for this device, for all devices presented in the main text and from additional devices. 

\begin{figure}
    \centering
    \includegraphics[width=\columnwidth]{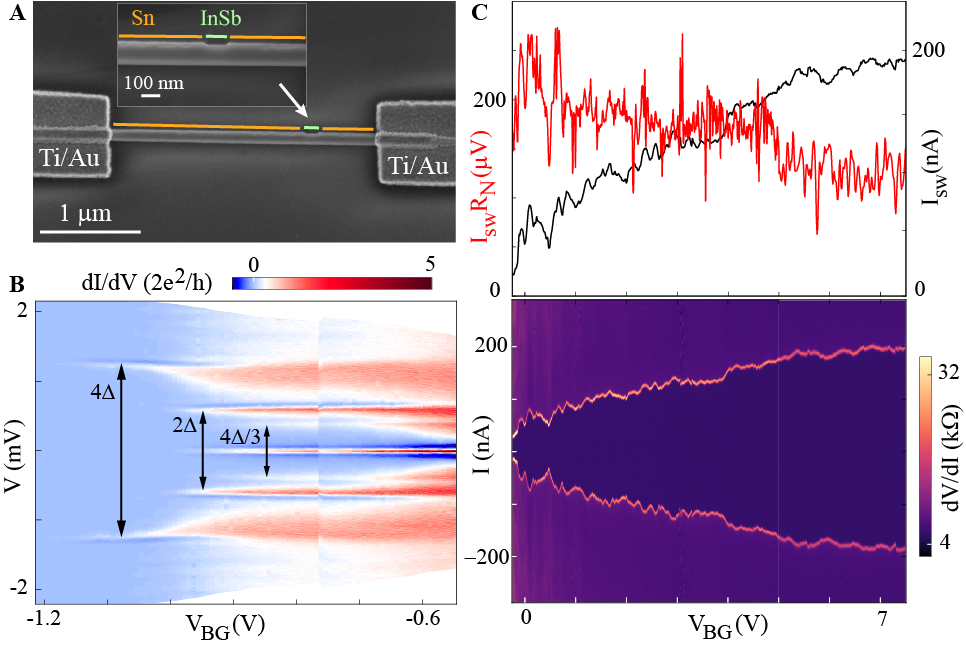}
    \caption{(A) SEM of Device B (S-S device). Inset zooms in the shadow junction where Sn shell is visible. (B) Differential conductance as a function of source-drain voltage bias $V$ and back gate voltage, V$_\textrm{BG}$. The double arrows mark resonances $4\Delta$, $4\Delta/2$ and $4\Delta/3$. (C) Differential resistance as a function of current bias $I$ and V$_\textrm{BG}$ (bottom). Top panel shows extracted switching current I$_\textrm{sw}$ (black) and I$_\textrm{sw}$R$_\textrm{N}$ (red) as a function of back gate voltage.}
    \label{fig2}
\end{figure}

Next, we study superconductor-superconductor (S-S) devices with both ends of the nanowire covered by tin, and only a narrow break in the shell to define an InSb weak link (Fig.~\ref{fig2}A). For this we use a previously developed method of shadowing the Sn flux by criss-crossing nanowires~\cite{gazibegovic2017epitaxy}. We first study tunneling between two tin islands (Fig.~\ref{fig2}B). We observe a smooth nanowire pinch off void of accidental quantum dot states. Three finite-bias resonances are observed, marked 4$\Delta$, 4$\Delta$/2 and 4$\Delta$/3 in Fig.~\ref{fig2}B. This sequence is a manifestation of Multiple Andreev Reflection processes, which are characteristic of transparent S-S junctions. They correspond to 615~$\mu$eV $\pm$ 10~$\mu$eV which is somewhat smaller than the gap observed in the N-S device (Fig.~\ref{fig1}C). At V$_\textrm{BG}<-1$~V only the 4$\Delta$ resonance is observed. We interpret this as the superconducting tunneling regime. Because the S-S tunneling resonance is a peak in current, it appears as a peak-dip structure in conductance. 

The resonance at zero bias in Fig.~\ref{fig2}B is the Josephson supercurrent. This effect is best studied in the current-bias configuration (Fig.~\ref{fig2}C). The switching current (I$_\textrm{sw}$) from superconducting to normal state is a peak in differential resistance. I$_\textrm{sw}$ decays smoothly with more negative V$_\textrm{BG}$. The current-voltage characteristics are weakly hysteretic which is reflected in the asymmetry of I$_\textrm{sw}$ in positive and negative current bias. In magnetic field, the Josephson effect is observed up to 1.5~T and remains significant with sharp switching up to 0.5~T (see Supporting Materials). This significant field range is a positive development for schemes that require coupling and decoupling of topologically superconducting islands at finite magnetic field for Majorana fusion or braiding~\cite{vanHeckNJP2012,AasenPRX16}. Measurements on continuous-shell nanowires without shadow junctions yielded supercurrents in the range $20-30~\mu$A corresponding to the critical current density of $2\times10^6$~A/cm$^2$ (data in Supporting Materials). The extracted products I$_\textrm{sw}$R$_\textrm{N}$ (R$_\textrm{N}$ is the normal state resistance) are in the range $125-225~\mu$eV, which is significant, and of the same order of magnitude as the gap. 

\begin{figure}
    \centering
    \includegraphics[width=\columnwidth]{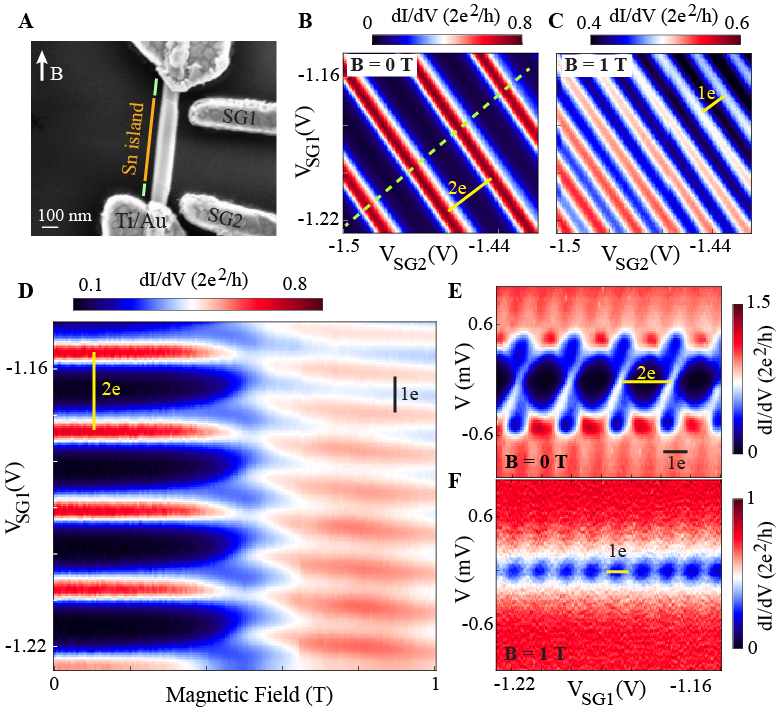}
    \caption{(A) SEM of device C showing the Sn island, two shadow junctions with bare InSb, side gates SG1 and SG2 and Ti/Au source-drain contacts which cover the outside Sn segments and suppress superconductivity there. (B) and (C): 2e and 1e tunneling conductance resonances measured at $V=0$ at B~$=0$~T and B~$=1$~T, respectively. (D) Magnetic field evolution of conductance along the dashed cut in panel (B). (E) and (F): $V$ vs gate spectroscopy at B~$=0$~T and B~$=1$~T, respectively.
    }
    \label{fig3}
\end{figure}

In Fig.~\ref{fig3} we present key findings on 2e charging of a tin island. The island is defined between two nanowire-shadow junctions in the N-S-N geometry (Fig.~\ref{fig3}A). At zero magnetic field, we observe a single family of Coulomb peak resonances consistent with charging the entire island (Fig.~\ref{fig3}B). At a finite magnetic field of 1~T, the frequency of Coulomb resonances doubles (Fig.~\ref{fig3}C,~\ref{fig3}D). We attribute data at zero field to 2e charging, and data at finite field to 1e charging. The transition from 2e to 1e is due to the superconducting gap or the lowest subgap state dropping in energy below the charging energy, which we estimate to be 0.3~meV (Figs.~\ref{fig3}E,~\ref{fig3}F). At finite magnetic field, it costs less energy to add electrons to the island one-by-one, while near zero field, due to hard gap superconductivity, it is advantageous to add electrons in pairs.

The two-electron charging effect is central for topological quantum computing because the states of a topological qubit are distinguished by even/odd island charge parity. If only 1e charging periodicity were observed, it would mean that despite a well-defined superconducting gap, electrons can be added to an island one at a time and the ability to distinguish the states of a topological qubit is scrambled. 1e periodicity is also detrimental for transmon qubits where single electron tunneling is a decoherence mechanism~\cite{PhysRevLett.121.157701}.

\begin{figure}
    \centering
    \includegraphics[width=\columnwidth]{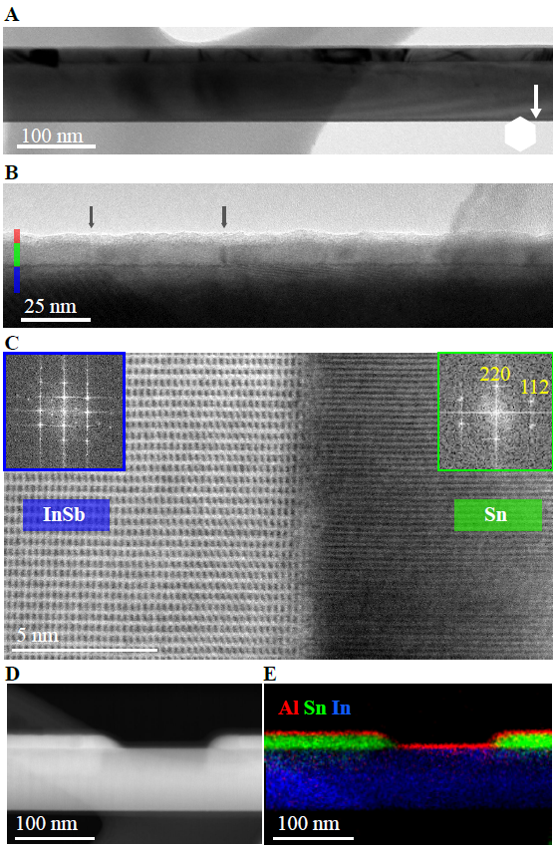}
    \caption{(A) Side-view TEM image along the $\langle$112$\rangle$ zone axis showing a homogeneously thin shell. (B) Higher magnification TEM of the $\mathrm{AlO_{x}}$ (red) - Sn (green) - InSb (blue) stack. The Sn grain boundaries are highlighted by arrows. (C) HR-STEM image of the Sn-InSb interface. The insets show Fourier transforms to the left and to the right of the interface. (D) High-angle annular dark-field STEM image of a shadow junction. (E) EDX elemental mapping of the shadow junction in (D). The Al-rich layer (red) corresponds to $\mathrm{AlO_{x}}$, oxygen not shown for clarity.}
    \label{fig4}
\end{figure}

Tin is an unusual material which has two different crystal phases with a phase transition at 13$\degree$C. The low-temperature $\alpha$-Sn has a diamond cubic lattice, while the high temperature $\beta$-Sn is tetragonal. The electronic properties of both phases are very different. $\alpha$-Sn is a semimetal which can also be a topological insulator in monolayer form~\cite{PhysRevLett.111.136804,PhysRevLett.118.146402,Zheng_2019_2dm}, whereas $\beta$-Sn is a metal with a superconducting transition temperature of 3.7~K.

For tin on InSb nanowires we assess the structural properties and elemental distribution using transmission electron microscopy (TEM). TEM images reveal a polycrystalline Sn shell of uniform thickness around the InSb nanowire (Fig.~\ref{fig4}A): the shell features grains of sizes $25\times25$~nm to $50\times60$~nm (see Supporting Materials). The Sn-InSb interface is abrupt and some Sn grains show epitaxial relationship with InSb. The high-resolution annular bright field-scanning TEM image (Fig.~\ref{fig4}C) shows a section of the interface where the $\{$111$\}$ planes of zincblende InSb are aligned with lattice planes of a Sn grain with a lattice distance of 2.04~$\textrm{\AA}$. This matches the $\{$220$\}$ interplanar distance of $\beta$-Sn. This grain is one of 13 analyzed along the same nanowire. 11 of the grains are identified as $\beta$-Sn from the fast Fourier transform analysis of the interplanar distances (Fig.~\ref{fig4}C, inset). Only two of those $\beta$-Sn grains show a preferential epitaxial relationship with InSb. In contrast,  $\alpha$-Sn is lattice matched to InSb and can grow epitaxially~\cite{farrow1981growth}. The predominantly $\beta$-Sn shell observed at room temperature by TEM is in agreement with superconductivity observed at low temperatures, suggesting that no phase transformation of Sn occurred upon device cooldown.

In addition to the uniform shell thickness, the nanowire shadow junctions used in S-S and N-S-N devices are sharp with Sn islands defined abruptly on each side of the junction (Fig.~\ref{fig4}C). Energy-dispersive x-ray spectroscopy (EDX) confirms that the Sn islands are isolated from each other and no interdiffusion between Sn and In is detected (Fig.~\ref{fig4}D). A uniform 3-nm-thick $\mathrm{AlO_{x}}$ passivation layer covers the entire nanowire. Oxidation at the Sn-InSb interfaces is not detected but cannot be fully excluded (see Methods).

Our results illustrate that neither defect-free epitaxial wire-shell interfaces nor single vacuum cycle growth of nanowire and shell are crucial requirements for the demonstration of hard gap, field-resilient superconductivity and 2e charging. We conclude and discuss in the Methods section that the key components in attaining robust induced superconductivity are (1) removal of InSb native oxide using atomic hydrogen prior to Sn growth, followed by (2) liquid nitrogen cooling of the nanowires during metal evaporation to produce a homogeneous ultrathin shell and (3) immediate passivation of the wire-shell hybrid with a stable dielectric.

We anticipate follow-ups of this work to be further research into the formation of Sn and InSb interfaces, and experiments in the Majorana geometry in search for robust signatures of topological superconductivity. Going forward, Sn/InSb nanowire system can be studied for applications in transmon and topological qubits. Beyond Sn, other metals can be tried as replacements of Al in search for decoherence free qubit materials~\cite{carrad2019shadow, bjergfelt2019superconducting, gusken2017mbe}. Interface requirements may increase for full topological qubit devices for which the dominant decoherence mechanisms are yet to be established experimentally.

\subsection*{Methods}

\textbf{Nanowire growth.} InSb nanowires are grown using the vapor-liquid-solid technique in a horizontal metal-organic vapor phase epitaxy reactor. The first nanowires used in this work are stemless InSb nanowires with flakes as shadow objects (Fig.~\ref{fig1}A)~\cite{badawy2019high, gazibegovic2019bottom}. Both InSb nanowires and flakes are grown on an InSb (111)B substrate with a selective-area mask and gold as catalyst.
The second type of nanowires, shown in Figs.~\ref{fig2}A and \ref{fig3}B are shadowed by other nanowires~\cite{gazibegovic2017epitaxy}. The InP (100) substrates are etched to expose the two $\{$111$\}$B facets, on which gold particles are deposited with an offset on the two opposing facets of a trench. Nanowires grow towards each other, such that the front wire shadows the back wire. InSb wires are grown on InP stems.

\textbf{Sn shell growth.} After transit in air, nanowire chips as grown are loaded into vacuum for subsequent growth of Sn shells. The chips are gallium bonded to molybdenum blocks. Atomic hydrogen cleaning is performed at 380$\degree$C (thermocouple temperature) for 30 minutes, at an operating pressure of $~5\times10^{-6}$~Torr consisting primarily of hydrogen ambient. Once cleaned, the samples are transferred in-vacuo to an ultra-high vacuum chamber dedicated for metal evaporation (base pressure $<5\times10^{-11}$~Torr). Here, the nanowire samples are cooled to 85~K (-188$^{\circ}$C) for 2 hours, prior to tin evaporation. 15-nm-thick tin is then evaporated from an effusion cell at a growth rate of 7.5~nm/hr and an evaporation angle close to 60$^{\circ}$ from sample normal. This shallow evaporation angle aids in-situ formation of Sn islands with nanowire or flake shadows. After Sn evaporation, while the sample is still expected to be at cryogenic temperatures (due to the thermal mass of the molybdenum block), a 3-nm-thick shell of $\mathrm{AlO_{x}}$ is electron-beam evaporated onto the nanowire sample, at normal incidence. The samples are then allowed to warm up to room temperature in vacuum.

\textbf{Device fabrication and measurements.} Device fabrication is similar to previous work on wires with epitaxial Al film~\cite{gazibegovic2017epitaxy}. Wires are transferred onto doped and thermally oxidized Si substrates using a micromanipulator under an optical microscope. Contacts and gates are patterned by electron-beam lithography by curing the resist at room temperature in vacuum to avoid nanowire heating and potential interdiffusion of Sn and In. Ar ion milling is performed to remove the $\mathrm{AlO_{x}}$ layer before evaporating 10/150 nm of Ti/Au. Measurements are performed in a dilution refrigerator with a 30~mK base temperature using a combination of direct current and lock-in techniques. All voltage bias data are two-terminal measurements. A series resistance of $\approx5k\Omega$ due to measurement setup was taken into account in calculating conductance in all figures as well as renormalizing $V$ axis in Figs~\ref{fig2}B,~\ref{figS1},~\ref{figS2}A. TEM studies were performed using a probe corrected microscope operated at 200~kV, equipped with a 100~mm$^2$ EDS detector.

\textbf{Acknowledgements.} Work supported by NSF PIRE-1743717, ANR HYBRID (ANR-17-PIRE-0001) and the Thomas Jefferson Fund. Solliance and the Dutch province of Noord Brabant are acknowledged for funding the TEM facility. CJP acknowledges Vannevar Bush Faculty Fellowship for characterization support at UCSB.

\bibliographystyle{apsrev4-1}
\bibliography{Ref.bib}

\pagebreak
\widetext
\clearpage

\begin{center}
\textbf{\large Supporting Materials}
\end{center}
\setcounter{figure}{0}
\setcounter{page}{1}
\renewcommand{\thefigure}{S\arabic{figure}}

\begin{figure*}[h]
    \centering
    \includegraphics[width=0.8\textwidth]{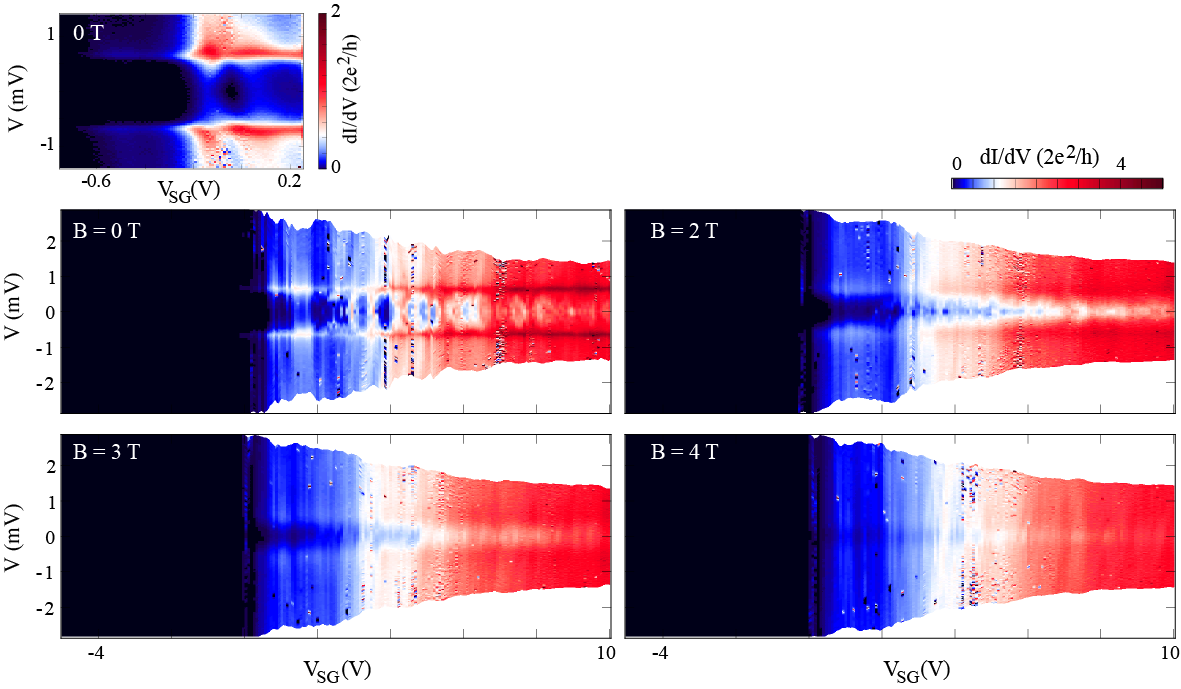}
    \caption{Additional gate dependence data for flake-shadowed device A studied in Fig.~\ref{fig1}. Zero-field hard gap regime displays the presence of a quantum dot in the vicinity of a tunnel barrier. Soft gap is observed for higher magnetic fields up to 4T. V$_\textrm{BG}=7.5$ V.
    }
    \label{figS1}
\end{figure*}

\begin{figure*}[h]
    \centering
    \includegraphics[width=0.6\textwidth]{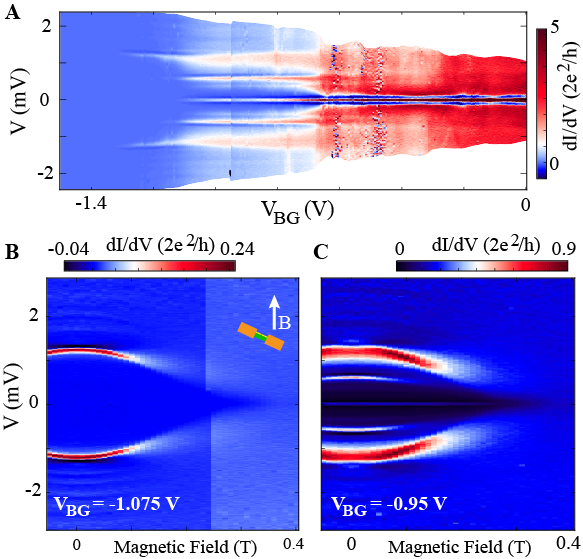}
    \caption{Additional data for device B studied in Fig.~\ref{fig2}. (A) Differential conductance as a function of bias voltage and back gate voltage in the extended gate range compared with Fig.\ref{fig2}. (B) and (C) show magnetic field dependence at V$_\textrm{BG}=-1.075$ V, and V$_\textrm{BG}=-0.95$ V, respectively. This device is at $\approx60^\circ$ angle with magnetic field and as a result the gap closes at B $\approx0.4$ T, a lower field  than in devices that were aligned parallel to the field, e.g. Fig.~\ref{fig1}.
    }
    \label{figS2}
\end{figure*}

\clearpage

\begin{figure*}[h]
    \centering
    \includegraphics[width=0.8\textwidth]{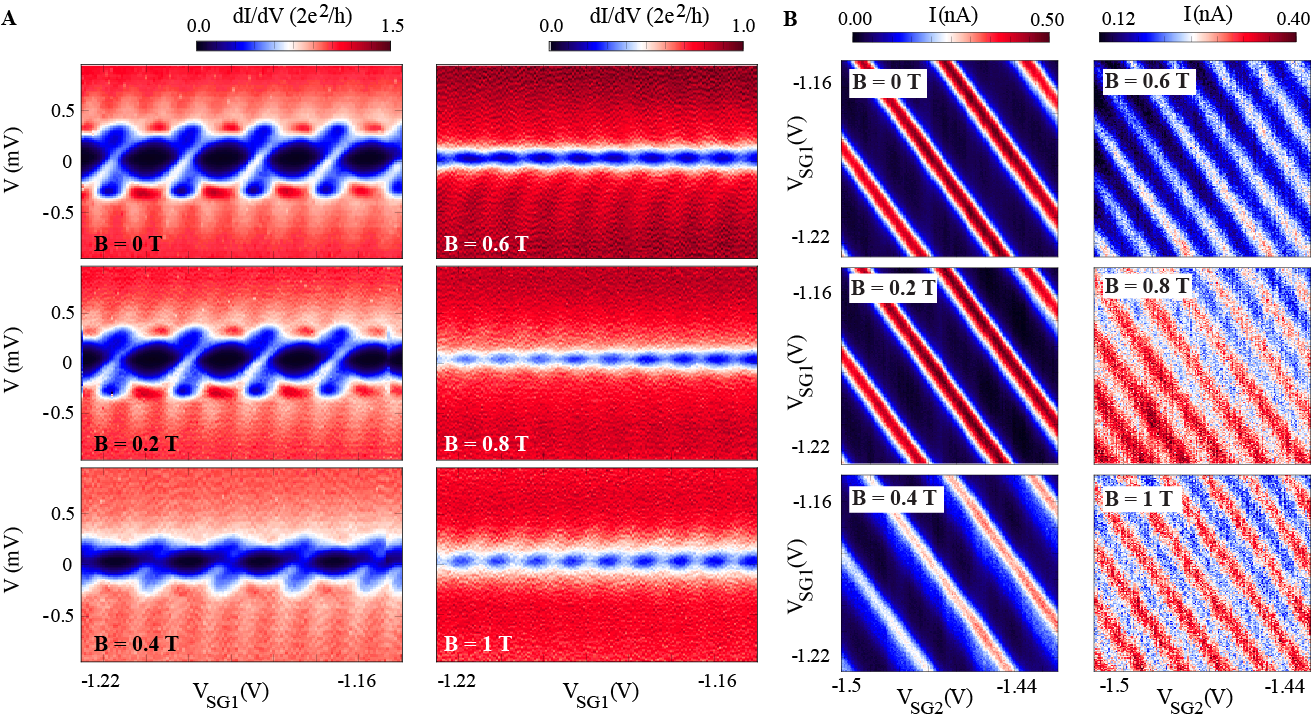}
     \caption{Additional data for device C studied in Fig.~\ref{fig3}. (A) Bias vs a combination of SG1 and SG2 at different magnetic fields, in the same regime as Fig.~\ref{fig3}E. (B) Gate dependent DC current measured at 10~$\mu$V bias. The gates are scanned in the same regime as Fig.~\ref{fig3}B}
    \label{figS3}
\end{figure*}

\begin{figure*}[t!]
    \centering
    \includegraphics[width=0.8\textwidth]{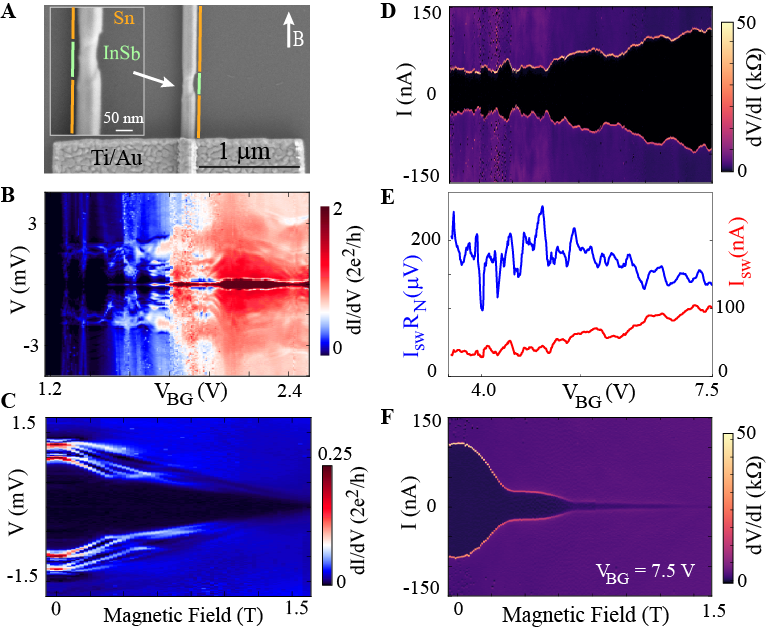}
    \caption{(A) SEM image of device D with a nanowire shadow defined break in the Sn shell. The electron beam resist was baked to 175$\degree$ in the process of making this device. Magnetic field is applied parallel to this device. (B) Differential conductance as a function of bias voltage and back gate voltage shows evolution from the supercurrent regime at more positive back gate to the quantum dot regime for more negative back gate. $V$ axis is as measured and not renormalized due to series resistance. (C) Andreev bound states evolution in parallel magnetic fields at V$_\textrm{BG}=1.4$~V. The resonances never reach zero bias due to hard gap. However, at $\textrm{B}=0.3-0.4$~T, the finite bias resonances exhibit kinks. These are the points where Andreev bound states cross zero chemical potential and undergo a ground state quantum phase transition. Resonances are shifted by $\pm\Delta(\textrm{B})$ in $\pm$~bias~\cite{SuPRL18}. In an N-S device a zero-bias peak would have been observed instead of finite-bias kinks. Gap remains open t 1.5~T in this device.
    (D) Current bias measurement in the more positive back gate regime showing the gate evolution of supercurrent. (E) Extracted $\textrm{I}_\textrm{sw}\textrm{R}_\textrm{N}$ (blue) and switching current (red) as a function of back gate voltage. (F) Magnetic field dependence of critical current.
    Josephson effect persists up to 1.5~T. This is consistent with the magnetic field decay of the induced gap (panel C). Previously, rapid decay of supercurrent in InSb nanowires was reported on the scale of 100~mT~\cite{ZuoPRL17}. It was explained in the context of interference of supercurrent carried by multiple occupied subbands. Here, supercurrent remains significant up to 0.5~T. This can be due to shorter junctions studied, enhanced screening of magnetic field by the Meissner effect in the Sn contacts or due to a lower subband occupation.
    }
    \label{figS4}
\end{figure*}

\clearpage

\begin{figure*}
    \centering
    \includegraphics[width=0.7\textwidth]{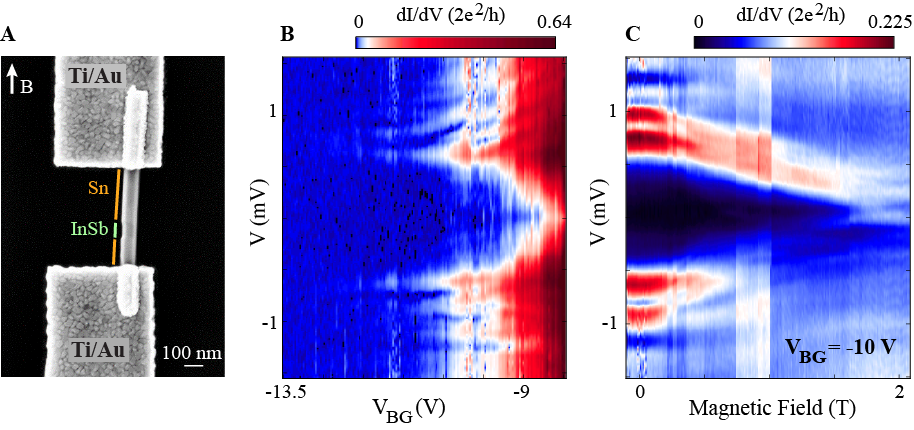}
    \caption{(A) SEM image of device E, a single shadow S-S device. The nanowire is aligned parallel with magnetic field. (B) Differential conductance as a function of bias and back gate voltages. $V$ axis is as measured and not renormalized due to series resistance. (C) Magnetic field evolution of conductance at V$_\textrm{BG}=-10$~V. Near $\textrm{B}=1.75$~T a zero bias conductance peak emerges from coalescence of two higher bias resonances. At these high magnetic fields the induced gap is soft, allowing for conductance at low bias, including the zero bias. The zero-bias peak is approximately 0.1 $2e^2/h$. We attribute this peak to a trivial zero-bias crossing by subgap Andreev states.}
    \label{figS5}
\end{figure*}

\begin{figure*}
    \centering
    \includegraphics[width=0.8\textwidth]{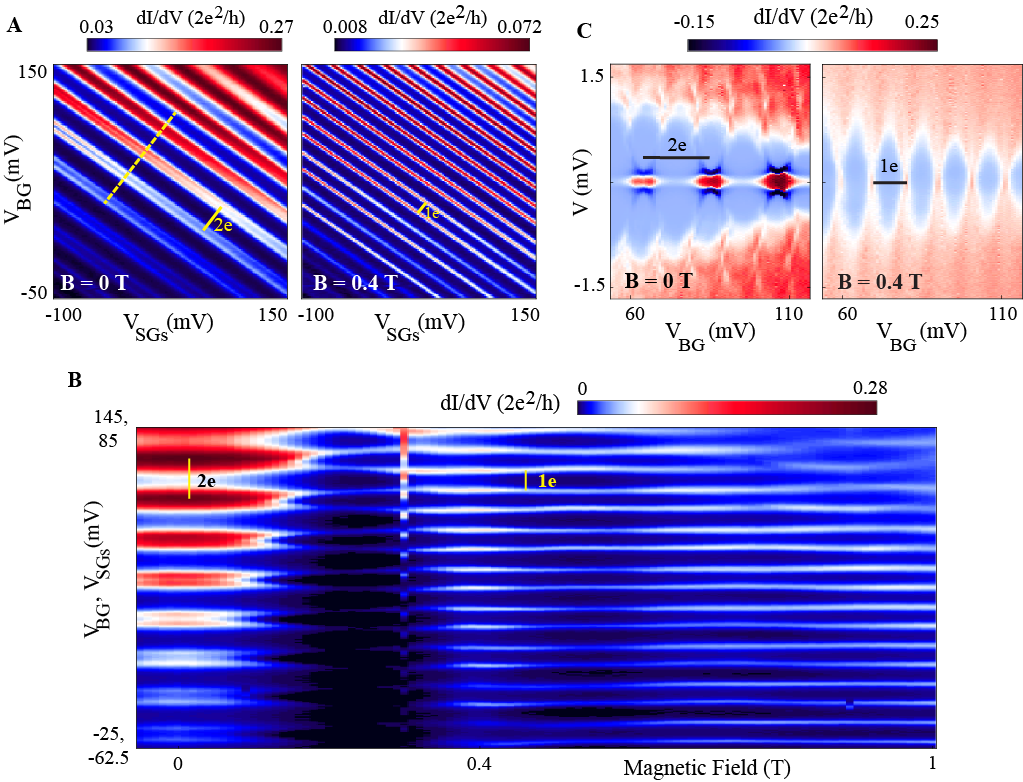}
    \caption{Device F is a two-shadow device of similar geometry to device C in Fig.~\ref{fig3}, however with a short Sn island of order 100 nm. For this device like device D the electron beam resist was baked at 175$\degree$ in the process of making side gates and contacts. Like device C, device F has three gates: two side gates aligned with the shadow junctions and a back gate. In contrast to device C which is in the N-S-N configuration, device F is in the S-S-S configuration with the center Sn island having superconducting leads due to the Sn shell. Nanowire F is at a 30$\degree$ angle with magnetic field. (A) Zero-bias conductance showing a 2e-periodic pattern of resonances at zero field and an 1e-periodic pattern at B~$=0.4$~T. Note that some resonances appear split. This is due to a subgap state that is lower in energy than the charging energy~\cite{Shen2018}. (B) Magnetic field evolution showing a transition from 2e-periodic to 1e-periodic pattern at finite field. (C) Differential conductance as a function of bias voltage and a combination of back gate and side gate voltages along the yellow dashed line in panel A. On the left, a zero-field scan reveals that the zero-bias conductance resonances are due to supercurrent through the S-S-S device which here manifest as zero-bias conductance peaks due to a voltage-bias measurement ~\cite{Veen2018}. At finite field (right) a pattern of Coulomb diamonds is observed and no supercurrent is observed.}
    \label{figS6}
\end{figure*}

\begin{figure*}
    \centering
    \includegraphics[width=0.8\textwidth]{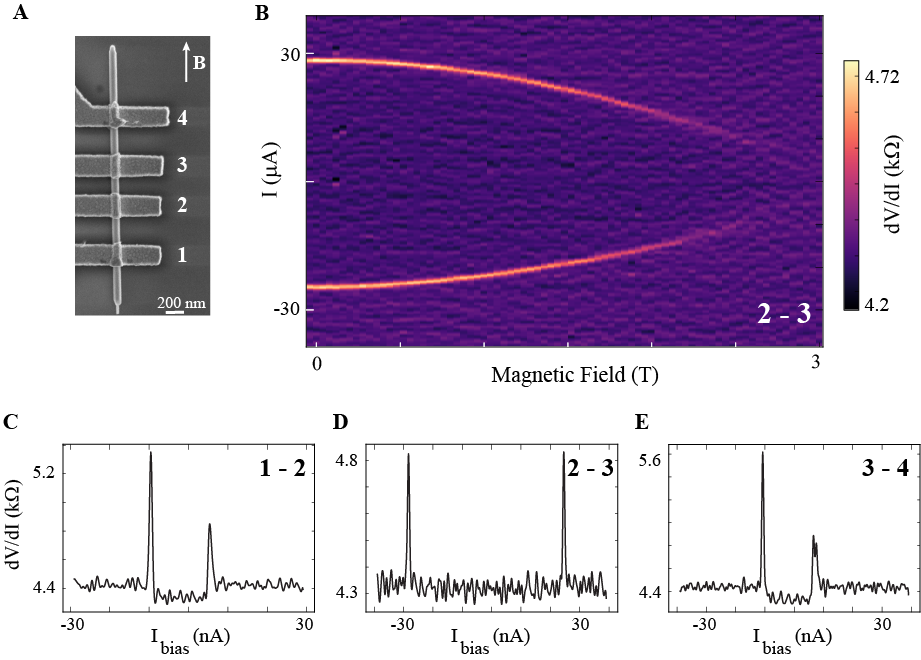}
    \caption{Device G has an uninterrupted Sn shell on an InSb nanowire without any shadows. The nanowire is grown on an InP stem. (A) SEM image showing 4 Ti/Au contacts labeled 1 to 4 noting different configurations for 2-terminal current bias measurements. (B) 2-terminal differential resistance as a function of bias current and magnetic field. Supercurrent persists up to 3 T, magnetic field is aligned parallel with the nanowire. (C-E) Differential resistance at zero magnetic field from different 2 terminal configurations. A variation in critical currents is observed along the shell with critical current being the highest in the central region of the nanowire. One possible explanation is the presence of grains of $\alpha$-Sn which are not superconducting, with a random distribution along the nanowire.
    }
    \label{figS7}
\end{figure*}
\clearpage

\begin{figure*}
    \centering
    \includegraphics[width=0.8\textwidth]{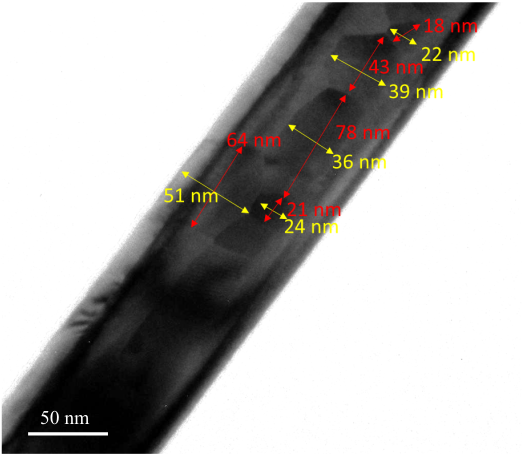}
    \caption{Side view bright field TEM image acquired along the $\langle$110$\rangle$ zone axis showing multiple Sn grains in the shell. Arrows and numbers are used to indicate the sizes of grains.}
    \label{figS8}
\end{figure*}
\clearpage

\begin{table}[H]
\begin{ruledtabular}
\begin{tabular}{ c c c c c c c c c }
Grain & $d_{hkl}$ (exp) & hkl $\beta$ & $d_{hkl}$ $\beta$ (lit) & \% deviation from lit $\beta$ & hkl $\alpha$  & $d_{hkl}$ $\alpha$ (lit) & \% deviation from lit $\alpha$ & epitaxy\\
\hline
A & 0.198 & 211 & 0.2010 & -1.2\% & 311 & 0.1956 & +1.5\% & \\
C & 0.209 & 220 & 0.2065 & +1.0\% & 311 & 0.1956 & +6.7\% & YES\\
D & 0.203 & 220 & 0.2065 & -1.4\% & 311  & 0.1956 & +4.1\% &\\
 & & 211 & 0.2010 & +1.3\% & - & - & - &\\
E & 0.206 & 220 & 0.2065 & -0.4\% & 311 & 0.1956 & +5.1\% &\\
F & 0.205 & 220 & 0.2065 & -0.5\% & 311 & 0.1956 & +5.0\% &\\
G & 0.274 & 101 & 0.2772 & -1.3\% & 211 & 0.264 & +3.7\% &\\
I & 0.282 & 101 & 0.2772 & +1.7\% & 211 & 0.264 & +6.7\% &\\
J & 0.277 & 101 & 0.2772 & -0.0\% & 211 & 0.264 & +5.0\% &\\
K & 0.203 & 220 & 0.2065 & -1.8\% & 311 & 0.1956 & +3.7\% &\\
L & 0.267 & 101 & 0.2772 & -3.7\% & 211 & 0.264 & +1.1\% &\\
M & 0.280 & 101 & 0.2772 & +1.2\% & 211 & 0.264 & +6.2\% &\\
N & 0.287 & 200 & 0.2920 & -0.8\% & - & - & - &\\
R & 0.204 & 220 & 0.2065 & -1.2\% & - & - & - &\\
  & 0.149 & 112 & 0.1472 & +1.3\% & - & - & - &\\
\end{tabular}
\end{ruledtabular}

    Table S1. Phase identification based on lattice spacings dhkl of 13 Sn grains imaged using high resolution TEM. All dhkl values are determined from Fast Fourier Transform patterns constructed from the HRTEM images. All patterns were calibrated by InSb lattice spacings present in the same images. The 211$\alpha$ spacing is not allowed based on crystal symmetry, but can appear in HRTEM images. The experimental inaccuracy in the dhkl values is estimated to be 2.0 percent considering the limited number of pixels in the FFT patterns. Based on this criterion, apart from grains A and L all grains can be assigned to the $\beta$-Sn phase. Grain R is presented in Fig.~\ref{fig4}C. 
\end{table}

\begin{figure*}[h]
    \centering
    \includegraphics[width=0.8\textwidth]{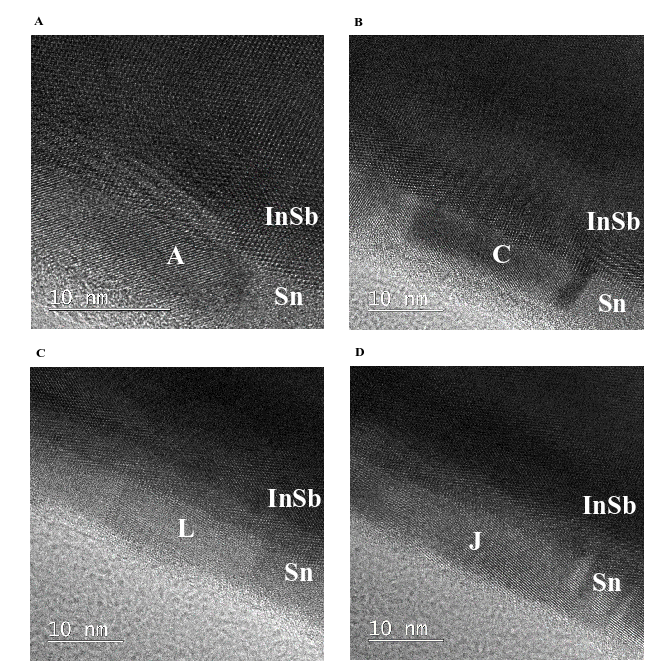}
    \caption{Side view TEM images of the core-shell interface, displaying four of the Sn grains listed in Table S1. (A) Grain A, which cannot unambiguously be assigned to either $\alpha$-Sn or $\beta$-Sn. (B) Grain C, $\beta$-Sn grain epitaxially related to the InSb lattice. (C) Grain L, $\alpha$-Sn. (D) Grain J, $\beta$-Sn.
    }
    \label{figS9}
\end{figure*}

\end{document}